\documentclass[11pt]{article}
\usepackage{graphicx,latexsym}
\usepackage{amsmath}
\usepackage{amsthm}
\usepackage{amssymb}
\usepackage{colortbl}
\usepackage{subcaption}
\usepackage{array}
\usepackage{tabularx}
\usepackage{stmaryrd}
\newcommand{\xvec}{{\bf x}}
\newcommand{\uvec}{{\bf u}}

\newcommand{\zerovec}{{\bf 0}}

\newtheorem{proposition}{Proposition}

\newtheorem{theorem}{Theorem}

\newtheorem{remark}{Remark}

\usepackage{epstopdf}
\begin{document}

\title{On the Observability and Controllability of Leaky-ReLU Networks}
\author{Liangjie Sun$^1$$^*$, Wai-Ki Ching$^2$, Shun-ichi Azuma$^3$, Tatsuya Akutsu$^4$}
\date{
   $^1$Institute for Life and Medical Sciences, Kyoto University, Kyoto 606-8507, Japan\\
   $^2$Department of Mathematics, The University of Hong Kong, Pokfulam Road, Hong Kong\\
   $^3$Graduate School of Informatics, Kyoto University, Kyoto 606-8501, Japan\\
   $^4$Bioinformatics Center, Institute for Chemical Research, Kyoto University, Kyoto 611-0011, Japan\\
   $^*$sunliangjie60@gmail.com}

\maketitle
\begin{abstract}
This paper studies minimum-node observability and controllability of Leaky rectified linear unit (Leaky-ReLU) networks under degree constraints. The objective is to characterize how many state nodes must be measured or
directly actuated to determine the initial state from a finite output sequence
or to steer the network between arbitrary states within a finite horizon.
For observability, a graph-theoretic analysis yields a class-wide upper bound
on the minimum number of observation nodes. We construct a family of networks
attaining this bound, thereby determining the exact worst-case minimum number
of observation nodes. We also construct networks that are observable from a
single node over a finite horizon, establishing the exact best-case value of
one. By establishing an observability--controllability duality under the
corresponding degree constraints, we obtain analogous exact best- and
worst-case results for the minimum number of control nodes.
A comparison with ReLU networks shows how replacing the zero negative slope
with a nonzero slope changes the observation-node requirement. More
generally, the observability arguments require only injectivity of the
activation function, whereas the controllability results extend to bijective
activation functions.
\end{abstract}
\section{Introduction}
In large-scale dynamical networks, measuring or directly actuating every state node is often impractical \cite{Olshevsky14,li22}. Sensors and external inputs can therefore be deployed at only a limited number of nodes, as commonly encountered in engineering and biological networks \cite{terasaki22,zhang25,zhang25ge,ben25}. This motivates the minimum-node observability and controllability problems: what are the minimum numbers of observation and control nodes required, respectively, to determine the initial state from a finite output sequence and to achieve a prescribed state transition within a finite horizon? General lower and upper bounds delimit the possible numbers of observation or control nodes required across the network class, while explicit constructions show whether these bounds can be attained.

Minimum-node observability and controllability have been studied for both finite- and continuous-state dynamical networks. In the finite-state setting, Boolean networks (BNs) have received particular attention. The average minimum number of driver nodes over a finite horizon was analyzed for random BNs with bounded in-degree \cite{hou16}. For conjunctive BNs, graph-theoretic methods were developed for minimum observability and controllability \cite{weiss18ob,weiss18co}. Graph-based methods were also used to study robust minimal strong reconstructibility of more general Boolean control networks \cite{li23robust}, while semi-tensor-product-based algebraic methods were developed for minimum observability of general BNs \cite{liu22}. Related minimum-node control problems have also been studied under structural controllability, pinning control, and stabilization frameworks \cite{zhu23,wang24pin,zhu24}. More recently, class-level lower and upper bounds on the minimum numbers of observation and control nodes have been derived for BNs in \cite{sun24} and \cite{sun26}, respectively.

In the continuous-state setting, related sensor- and actuator-selection problems have been studied under structural functional observability and generic state-and-input observability \cite{zhang25,cheng25}, as well as structural output controllability and structural controllability \cite{zhang25ge,guo21}. Pinning-control methods have also been developed to synchronize dynamical networks through selected nodes and to identify effective pinned-node sets \cite{chen17,liu25pin}. Separately, observability and controllability of prescribed piecewise-affine and hybrid systems have been investigated using mixed-integer formulations \cite{Bemporad2000}. However, existing results largely concern prescribed systems or fixed network structures and do not directly establish class-level lower and upper bounds on the minimum numbers of observation and control nodes for activation-based continuous-state networks. They also leave open how these bounds are shaped by the activation function and the network interconnection structure.

The Leaky rectified linear unit (Leaky-ReLU) provides a particularly informative setting for addressing these questions. For a leakage parameter $0<\alpha<1$, Leaky-ReLU \cite{maas2013} has unit slope on the nonnegative half-line and slope $\alpha$ on the negative half-line. It is therefore continuous, piecewise linear, and bijective on $\mathbb R$. When extended to $\alpha\in[0,1]$, this activation family interpolates between ReLU at $\alpha=0$ and the identity map at $\alpha=1$. In our previous work, we derived upper and lower bounds on the minimum number of observation nodes for ReLU networks \cite{sun2026}.
Unlike ReLU, Leaky-ReLU does not collapse all negative preactivations to zero or restrict the activation output to the nonnegative half-line. At the same time, Leaky-ReLU networks retain state-dependent piecewise-affine dynamics. They thus provide a natural model for determining which limitations on observability and controllability arise from the noninvertibility of ReLU and which persist because of the network interconnection structure.

The main contributions of this paper are summarized as follows.
\begin{itemize}
\item[(i)]
For any Leaky-ReLU networks with maximum indegree $K$,
we derive a class-wide upper bound on the minimum number of observation
nodes. We then construct a family of networks for which this number cannot
be reduced below the derived bound. Hence, the upper bound is attained and
the exact worst-case observation-node requirement is determined. By
establishing an observability--controllability duality for the considered
network class, we obtain the corresponding control-node result.
\item[(ii)]
We identify structured Leaky-ReLU networks that are observable from a
single node over a finite horizon. Since at least one observation node is
necessary, the exact best-case observation-node requirement is one. We
similarly construct networks that are controllable from a single control
node, establishing the corresponding exact best-case controllability result.
\item[(iii)]
We compare the resulting observation-node bounds with those for ReLU
networks under the same network-size and degree constraints. The comparison
shows how replacing the zero negative slope of ReLU by a nonzero slope
changes the node requirements. More generally, the observability arguments
extend to networks with componentwise injective activation functions, while
the controllability arguments extend to networks with componentwise
bijective activation functions.
\end{itemize}

\section{Problem Formulation}\label{section2}
Throughout the paper, let $\mathbb{R}$ and $\mathbb{R}_{>0}$ denote the sets of real and
positive real numbers, respectively. For integers $a\leq b$, let
$\llbracket a,b\rrbracket:=\{a,a+1,\ldots,b\}$.

Consider the following Leaky-ReLU network, whose maximum indegree is $K$; that is, the update function of each node depends on at most $K$ current-state variables, and at least one depends on exactly $K$ such variables:
\begin{eqnarray*}
x_{i}(t+1) & =  & \rho\left(\sum_{j=1}^{d^{-}(x_i)}a_{i,j}x_{i_{j}}(t)+b_{i}\right),\quad \forall i\in\llbracket 1,n\rrbracket,
\end{eqnarray*}
where $x_{i}(t)\in\mathbb{R}$ is the state of node $x_{i}$, $a_{i,j}\in \mathbb{R}_{>0}$, $b_{i}\in\mathbb{R}$, $d^{-}(x_i)$ is the indegree of node $x_{i}$, with $d^{-}(x_i)\leq K$ for all $i$ and $\max_i d^{-}(x_i)=K$, $i_j\in\llbracket 1,n\rrbracket$ denotes the index of the $j$-th in-neighbor of node $x_{i}$, and $\rho(s)=\max(s,\alpha s),~0<\alpha<1$ is the Leaky-ReLU function. Let $\xvec(t)=(x_{1}(t),\ldots,x_{n}(t))$.

We next introduce some basic graph-theoretic definitions. Let $G=(V,E)$ be the underlying directed graph of the Leaky-ReLU network, where $V=\{x_1,x_2,\ldots,x_n\}$. A node $x_{i_{j}}$ is an in-neighbor of node $x_i$ if and only if $(x_{i_{j}},x_i)\in E$.

A set of nodes $S\subseteq V$ is called \emph{strongly connected}
if, for any two distinct nodes $x_{p},x_{q}\in S$, there exists a directed path from $x_{p}$ to $x_{q}$.
A \emph{strongly connected component} (SCC) is a maximal strongly connected subset of $V$.
The subgraph induced by an SCC is also called an SCC when no confusion arises.

\begin{figure}[ht]
\centering
\includegraphics[width=10cm]{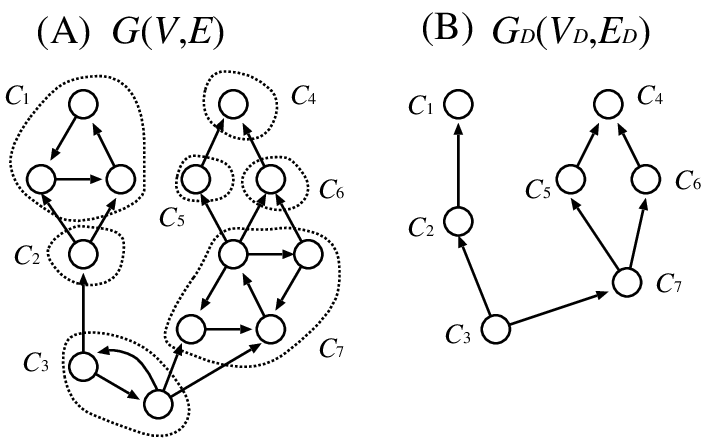}
\caption{Example of decomposition based on SCCs.
(A) Original graph $G=(V,E)$.
(B) Directed acyclic graph $G_D=(V_E,E_D)$ constructed from strongly connected components. Here, $C_1$ and $C_4$ are the top SCCs, so $N=2$.
}
\label{fig:scc}
\end{figure}

By contracting each SCC into a single node, we obtain the resulting graph, denoted by $G_D=(V_D,E_D)$, which is a directed acyclic graph (DAG), as illustrated in Fig.~\ref{fig:scc}. Furthermore, for two SCCs $C_{1}$ and $C_{2}$ in $G_{D}$, we say that
$C_1$ is a \emph{predecessor} of $C_2$ if there exists
a directed edge from $C_2$ to $C_1$.\footnote{This definition is different
from (inverse of) the standard definition of a predecessor,}
Note that an SCC may have multiple predecessors.
We also say that $C_1$ is an \emph{ancestor} of $C_2$ if
$C_1$ is a predecessor of $C_2$, or
there exists a predecessor $C_3$ of $C_2$ such that $C_1$ is an ancestor of $C_3$. An SCC $C\in V_{D}$ is called a \emph{top SCC} if it has no outging edge in $G_{D}$. Let $N$ denote the number of top SCCs in $G_{D}$. Although the concept of a ``top SCC'' was employed in \cite{pequito15}, our analysis differs substantially from the existing one.

We consider the following two problems.
\begin{itemize}
  \item \textbf{Bounds on the minimum number of observation nodes.} Suppose that $m$ nodes, indexed by $j_1,j_2,\ldots,j_m$, are designated as observation nodes. For each $l\in\llbracket 1,m\rrbracket$, state $x_{j_l}(t)$ can be directly available for measurement. The Leaky-ReLU network is said to be observable with respect to these observation nodes if any two distinct initial states $\xvec(0)\neq \bar{\xvec}(0)$ can be distinguished by $x_{j_1}(t),x_{j_2}(t),\ldots,x_{j_m}(t),~t\ge 0$. We derive upper and lower bounds on the minimum number of observation nodes required to ensure observability.
  \item \textbf{Bounds on the minimum number of control nodes.} Suppose that $q$ nodes, indexed by $j_1,j_2,\ldots,j_q$, are designated as control nodes. For each $l\in\llbracket 1,q\rrbracket$, state $x_{j_{l}}(t)$ can be directly assigned by an external control input, that is, $x_{j_{l}}(t+1)=u_{j_{l}}(t)$, where $u_{j_{l}}(t)\in\mathbb{R}$ can be chosen arbitrarily. Let $\uvec(t)=(u_{j_{1}}(t),u_{j_{2}}(t),\ldots,u_{j_{q}}(t))$. The remaining nodes evolve according to the original network dynamics. The Leaky-ReLU network is said to be controllable with respect to these control nodes if, for any initial state $\xvec(0)$ and any target state $\xvec^\ast$, there exist a finite time $T\ge 0$ and a control sequence $\uvec(0),\uvec(1),\ldots,\uvec(T-1)$ such that $\xvec(T;\xvec(0),\uvec)=\xvec^\ast$, where $\xvec(T;\xvec(0),\uvec)$ denotes the state at time $T$ starting from the initial state $\xvec(0)$ under the control sequence $\uvec=(\uvec(0),\uvec(1),\ldots,\uvec(T-1))$. We derive upper and lower bounds on the minimum number of control nodes required to ensure controllability.
\end{itemize}
\section{Optimal Bounds for the Minimum Number of Observation Nodes}\label{sec:obs}
In this section, we derive upper and lower bounds on the minimum number of observation nodes required for observability.

We first consider Leaky-ReLU networks with maximum indegree $K=2$. In other words, each node has at most two incoming edges, and at least one node has exactly two incoming edges.

\begin{figure}[ht]
\centering
\includegraphics[width=8cm]{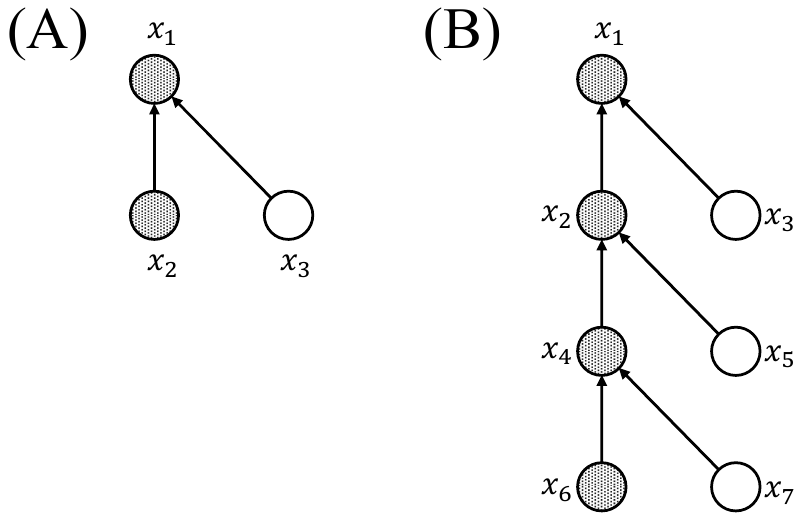}
\caption{Leaky-ReLU networks for explaining the basic idea.
By choosing gray nodes as observation nodes,
the system becomes observable.}
\label{fig:basic}
\end{figure}

To illustrate the basic idea, consider the three-node network with edge set $E=\{(x_2,x_1),(x_3,x_1)\}$, as shown in Fig.~\ref{fig:basic}(A). Suppose that $x_{1}$ and $x_{2}$ are selected as observation nodes. It follows from
\begin{eqnarray*}
x_3(t) = \frac{\rho^{-1}(x_1(t+1)) - a_{1,2} x_2(t) - b_1}{a_{1,3}},
\end{eqnarray*}
that initial state of the non-observation node $x_{3}$, and hence the entire initial state $\xvec(0)$ can be uniquely determined from observations $x_{1}(0),x_{1}(1)$, and $x_{2}(0)$.

The same idea can be extended to larger networks. Consider the seven-node network with edge set $E=\{(x_2,x_1),(x_3,x_1),(x_4,x_2),(x_5,x_2),(x_6,x_4),(x_7,x_4)\}$, as shown in Fig.~\ref{fig:basic}(B). By choosing nodes $x_{1},x_{2},x_{4}$, and $x_{6}$ as observation nodes, the initial state $\xvec(0)$ can be uniquely determined from $x_{1}(0), x_{1}(1), x_{2}(0), x_{2}(1), x_{4}(0), x_{4}(1)$, and $x_{6}(0)$.

For the above example, the required number of observation nodes can also be obtained from the SCC decomposition of $G=(V,E)$ in Fig.~\ref{fig:basic}(B). Here, $N=1$ and $n=7$, and thus the number of observation nodes is equal to $1+(7-1)/2=4$.

Based on this idea, we propose a greedy-type procedure for selecting observation nodes. Although the underlying idea is simple, its detailed implementation requires some care. We first classify the nodes $x_i,~i\in\llbracket 1,n\rrbracket$, into the following three types:
\begin{description}
\item [{\rm $\ell(x_i)=0$:}] $x_i$ is not yet processed,
\item [{\rm $\ell(x_i)=1$:}] $x_i$ is an observation node,
\item [{\rm $\ell(x_i)=2$:}] $x_i$ is non-observation node, but can be observed.
\end{description}
The procedure is based on the following observation.
\begin{itemize}
\item If $d^{-}(x_i)=1$, $(x_j,x_i) \in E$ and $x_i(t)$ can be observed,
then $x_j(t)$ can be observed.
\item If $d^{-}(x_i)=2$, $(x_j,x_i) \in E$, $(x_k,x_i) \in E$, and
$x_i(t)$ and $x_k(t)$ can be observed,
then $x_j(t)$ can be observed.
\item If $d^{-}(x_i)=2$, $(x_j,x_i) \in E$, $(x_k,x_i) \in E$, and
$x_i(t)$ can be observed,
then $x_j(t)$ and $x_k(t)$ can be observed by selecting
$x_j$ as an observation node.
\end{itemize}
Note that the number of observation nodes increases by 1
only in the last case, for which the number of observable nodes
increases by 2 (Observation (\$1)).

The following is the pseudo-code of the procedure to select
the set of observation nodes.
\begin{rm}
\begin{tabbing}
\quad \= \quad \= \quad \= \quad \= \quad \= \quad \= \quad \= \quad \= \kill
\> \> Procedure $SelObsNodesK2(G(V,E))$\\
\> \> \> {\bf for all} nodes $x_i$ {\bf do} $\ell(x_i) \leftarrow 0$;\\
\> \> \> {\bf for all} top SCC $C$ {\bf do} choose any $x_i \in C$;~$\ell(x_i) \leftarrow 1$;~~~~(\#1)\\
\> \> \> {\bf while} there exists an SCC $C$ such that $x_h\in C$ with $\ell(x_h)=0$ and either $C$ is a top SCC\\
\> \> \> or $C$ has an outgoing edge to some SCC $C'$ with $\ell(x_j)>0$ for all $x_{j}\in C'$ {\bf do}~~~~~(\#2)\\
\> \> \> \> choose such an SCC $C$;\\
\> \> \> \> {\bf while} there exists $x_h \in C$ with $\ell(x_h)=0$ {\bf do}\\
\> \> \> \> \> {\bf let} $(x_j,x_i)$ be an edge with $x_j \in C$ such that
$\ell(x_i)>0$ and $\ell(x_j)=0$;~~~~~~~~~(\#3)\\
\> \> \> \> \> {\bf if} $d^{-}(x_i)=1$ {\bf then} $\ell(x_j) \leftarrow 2$ {\bf else}\\
\> \> \> \> \> \> let $x_k$ be the node such that $(x_k,x_i) \in E$;\\
\> \> \> \> \> \> {\bf if} $\ell(x_k)>0$ {\bf then} $\ell(x_j) \leftarrow 2$ {\bf else} $\ell(x_j) \leftarrow 1$;  $\ell(x_k) \leftarrow 2$.
\end{tabbing}
\end{rm}

For the case of Fig.~\ref{fig:scc},
$C_1$ and $C_4$ will be selected at STEP (\#1) and
$C_2$, $C_5$, $C_6$, $C_7$, and $C_3$ may be selected at STEP (\#2) in this order.

\begin{theorem}
For any Leaky-ReLU network with maximum indegree $K=2$,
$SelObsNodesK2(G(V,E))$ returns, in polynomial time, a set of at most
$N+\left\lfloor \frac{n-N}{2}\right\rfloor$ observation nodes that ensures observability.
\label{thm:general-upper-k2}
\end{theorem}
(Proof) We first consider the case where $n-N$ is even. The procedure selects at most $N+\frac{n-N}{2}$ observation nodes.
In STEP (\#1), it selects $N$ nodes from the $N$ top SCCs. After that, each additional observation node makes two more nodes observable, from Observation (\$1).

We next consider the case where $n-N$ is odd, the above procedure also selects at most $N+\frac{n-N-1}{2}$ observation nodes. Indeed, after selecting $N+\frac{n-N-1}{2}$ nodes, the $n-1$ nodes become observable, and then the remaining node can always be observed.

Next, it is seen that the status of each node $x_i$ becomes
$\ell(x_i)=1$ or $\ell(x_i)=2$
after the termination of the procedure because
\begin{itemize}
\item for any SCC $C$ selected at STEP (\#2),
there exists at least one $x_j \in C$ with $(x_j,x_i) \in E$ and $\ell(x_i)>0$,
\item since SCC is strongly connected,
each node in SCC (except any top SCC consisting of a single node)
has an incoming node,
\item accordingly, a node $x_j$ can always be found at STEP (\#3).
\end{itemize}

Finally, it is straightforward to see that the procedure works in
polynomial time.
\qed

Note that Theorem~\ref{thm:general-upper-k2} does not claim that $SelObsNodesK2(G(V,E))$ finds a minimum-cardinality set of observation nodes. It only guarantees a set of at most $N+\left\lfloor \frac{n-N}{2}\right\rfloor$ observation nodes that ensures observability. This gives a general upper bound on the minimum number of observation nodes required for observability.

\begin{proposition}
There exists a Leaky-ReLU network with maximum indegree 2 that need
at least $N+\left\lfloor \frac{n-N}{2}\right\rfloor$ observation nodes.
\label{prop:worst-lower-k2}
\end{proposition}
(Proof) We construct the desired network by taking $N$ disjoint copies of the network shown in Fig.~\ref{fig:basic}(B), where the path lengths in these copies are chosen so that the total number of nodes is $n$.
Then, the proposition clearly holds.
\qed

In Proposition \ref{prop:worst-lower-k2}, we establish a worst-case lower bound. Here, ``worst-case'' means that although some networks may require fewer observation nodes, there exists a network in the considered class for which fewer than $N+\left\lfloor \frac{n-N}{2}\right\rfloor$ observation nodes cannot ensure observability.

\begin{remark}
The general upper bound on the minimum number of observation nodes matches the corresponding worst-case lower bound, showing that the upper bound is tight.
\end{remark}

\begin{figure}[ht]
\centering
\includegraphics[width=8cm]{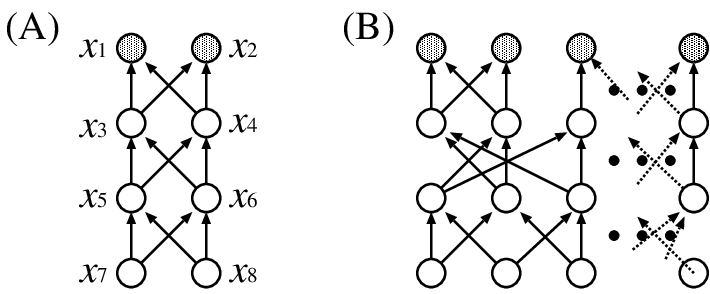}
\caption{Leaky-ReLU networks with a small number of observation nodes.}
\label{fig:best-upper}
\end{figure}

Next, we have the following result.
\begin{proposition}
There exists a Leaky-ReLU network with indegree 2 at every node that is observable from two observation nodes.
\label{prop:best-upper-k2}
\end{proposition}
(Proof)
Consider the network shown in Fig.~\ref{fig:best-upper}(A),
\begin{eqnarray*}
x_{2i-1}(t+1) & = & \rho(a_{2i-1,2(i+1)-1}x_{2(i+1)-1}(t) + a_{2i-1,2(i+1)}x_{2(i+1)}(t) + b_{2i-1}),\\
x_{2i}(t+1) & = & \rho(a_{2i,2(i+1)-1}x_{2(i+1)-1}(t) + a_{2i,2(i+1)}x_{2(i+1)}(t) + b_{2i}),
\end{eqnarray*}
where $i\in\llbracket 1,\frac{n}{2}\rrbracket$, $n$ is even, and
$\begin{vmatrix}
a_{2i-1,2(i+1)-1} & a_{2i-1,2(i+1)} \\
a_{2i,2(i+1)-1} & a_{2i,2(i+1)}
\end{vmatrix}\neq0$.
For $i=\frac{n}{2}$, the indices $2(i+1)-1$ and $2(i+1)$ are identified with 1 and 2, respectively. According to
\begin{eqnarray*}
a_{2i-1,2(i+1)-1}x_{2(i+1)-1}(t) + a_{2i-1,2(i+1)}x_{2(i+1)}(t) & = & \rho^{-1}(x_{2i-1}(t+1)) - b_{2i-1},\\
a_{2i,2(i+1)-1}x_{2(i+1)-1}(t) + a_{2i,2(i+1)}x_{2(i+1)}(t)& = & \rho^{-1}(x_{2i}(t+1)) - b_{2i},
\end{eqnarray*}
it follows that $x_{2(i+1)-1}(t)$ and $x_{2(i+1)}(t)$ can be uniquely determined by $x_{2i-1}(t+1)$ and $x_{2i}(t+1)$, since
\begin{eqnarray*}
\begin{bmatrix}
x_{2(i+1)-1}(t) \\
x_{2(i+1)}(t)
\end{bmatrix}=
\begin{bmatrix}
a_{2i-1,2(i+1)-1} &  a_{2i-1,2(i+1)}\\
a_{2i,2(i+1)-1} & a_{2i,2(i+1)}
\end{bmatrix}^{-1}
\begin{bmatrix}
\rho^{-1}(x_{2i-1}(t+1)) - b_{2i-1}\\
\rho^{-1}(x_{2i}(t+1)) - b_{2i}
\end{bmatrix}.
\end{eqnarray*}
Consequently, $x_3(0)$ and $x_4(0)$ can be recovered from $x_1(1)$ and $x_2(1)$. Repeated application of the above relation then recovers $x_5(0)$ and $x_6(0)$ from $x_1(2)$ and $x_2(2)$, and, more generally, $x_{2r+1}(0)$ and $x_{2r+2}(0)$ from $x_1(r)$ and $x_2(r)$, for $r\in\llbracket 1,\frac{n}{2}-1\rrbracket$. Thus, the entire initial state $\xvec(0)$ can be uniquely determined from the observations $x_1(t),x_2(t),~t\in\llbracket 0,\frac{n}{2}-1\rrbracket$.
\qed

\begin{remark}
The network shown in Fig.~\ref{fig:best-upper}(A) can be extended to
a layered network with $L$ nodes in each layer, as illustrated in Fig.~\ref{fig:best-upper}(B). For this extended network, the minimum number of observation nodes required for observability is $L$.

In this case, the set of edges connecting the $i$-th layer nodes
and the $(i+1)$-th layer nodes can be regarded as a bipartite graph,
where the top layer in Fig.~\ref{fig:best-upper}(B)
is regarded as the first layer. If this bipartite graph has a perfect matching,
then the determinant of the corresponding adjacency matrix
will not be zero for most parameters $a_{i,j}$.
Consequently, the values $x_h(t)$ for all nodes $h$ in the $(i+1)$-th layer can be uniquely determined from the values
$x_k(t+1)$ for all nodes $k$ in the $i$-th layer.
\end{remark}

\begin{proposition}
There exists a Leaky-ReLU network with maximum indegree $K=2$ that is observable from a single observation node.
\label{prop:best-case}
\end{proposition}
(Proof)
Construct a Leaky-ReLU network with maximum indegree $K=2$ as follows:
\begin{eqnarray*}
x_{1}(t+1) & = & \rho(a_{1,2}x_{2}(t)+b_{1}),\\
x_{2}(t+1) & = & \rho(a_{2,1}x_{1}(t)+a_{2,3}x_{3}(t)+b_{2}),\\
x_{3}(t+1) & = & \rho(a_{3,2}x_{2}(t)+a_{3,4}x_{4}(t)+b_{3}),\\
&\ldots&\\
x_{n-1}(t+1) & = & \rho(a_{n-1,n-2}x_{n-2}(t)+a_{n-1,n}x_{n}(t)+b_{n-1}),\\
x_{n}(t+1) & = & \rho(a_{n,n-1}x_{n-1}(t)+a_{n,1}x_{1}(t)+b_{n}).
\end{eqnarray*}
From the update equations,
$x_2(0)=\frac{\rho^{-1}(x_1(1))-b_{1}}{a_{1,2}}$,
and
\begin{eqnarray*}
x_i(0)=\frac{\rho^{-1}(x_{i-1}(1))-a_{i-1,i-2}x_{i-2}(1)-b_{i-1}}{a_{i-1,i}},\quad i\in\llbracket 3,n \rrbracket.
\end{eqnarray*}
By recursively applying these relations, $x_i(0)$ can be uniquely determined from $x_1(i-1),\ x_1(i-3),\ldots$, ending with $x_1(0)$ when $i$ is odd and with $x_1(1)$ when $i$ is even. Thus, the entire initial state $\xvec(0)$ can be uniquely determined from the observations $x_1(t),~t\in\llbracket 0,n-1 \rrbracket$.
\qed

In Proposition \ref{prop:best-case}, we establish a best-case upper bound. Here, ``best case'' means that although some networks may require more observation nodes, there exists a network in the considered class for which no more than one observation node is needed to ensure observability.

\begin{remark}
At least one observation node is necessary for observability, so 1 is a general lower bound on the minimum number of observation nodes. Proposition \ref{prop:best-case} constructs a network with maximum indegree $K=1$ that is observable from a single observation node. Thus, this lower bound is attained and is therefore tight.
\end{remark}

Now we extend the above results to $K>2$.

\begin{theorem}
For any Leaky-ReLU network with maximum indegree $K$, a set of at most $N+\left\lfloor\frac{(K-1)(n-N)}{K} \right\rfloor$
observation nodes that ensures observability.
\end{theorem}
(Proof)
Similar to $K=2$, we can see that when $d^{-}(x_{i})=K$, $(x_{i_{1}},x_{i})\in E, (x_{i_{2}},x_{i})\in E, \ldots, (x_{i_{K}},x_{i})\in E$, there are $K$ cases as follows:
\begin{itemize}
  \item If $x_{i}(t),x_{i_{1}}(t),\ldots,x_{i_{K-1}}(t)$ can be observed, then $x_{i_{K}}$ can be observed.
  \item If $x_{i}(t),x_{i_{1}}(t),\ldots,x_{i_{K-2}}(t)$ can be observed, then $x_{i_{K-1}},x_{i_{K}}$ can be observed by letting $x_{i_{K-1}}$ as an observation node.
  \item $\ldots$
  \item If $x_{i}(t)$ can be observed, then $x_{i_{1}}(t),\ldots,x_{i_{K}}$ can be observed by letting $x_{i_{1}},\ldots,x_{i_{K-1}}$ as observation nodes.
\end{itemize}
In the worst case, adding $K-1$ observation nodes makes only $K$ additional nodes observable. Therefore, the minimum number of observation nodes is at most $N+\left\lfloor\frac{(K-1)(n-N)}{K} \right\rfloor$.
\qed

\begin{figure}[ht]
\centering
\includegraphics[width=5cm]{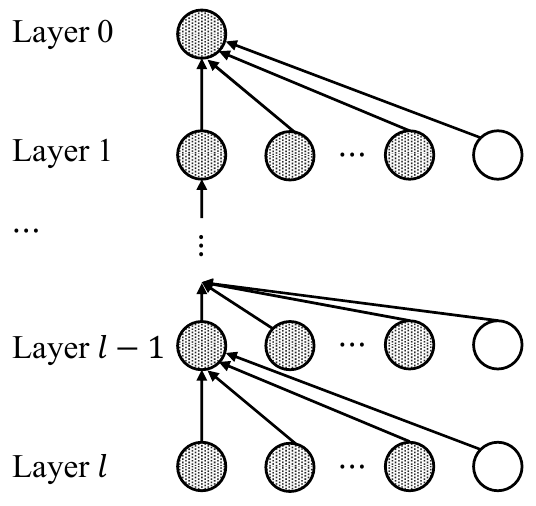}
\caption{Leaky-ReLU networks with maximum indegree $K$.}
\label{fig5}
\end{figure}

\begin{proposition}
There exists a Leaky-ReLU network with maximum indegree $K$ that need at least $N+\left\lfloor \frac{(K-1)(n-N)}{K} \right\rfloor$ observation nodes.
\end{proposition}
(Proof)
Construct a Leaky-ReLU network with maximum indegree $K$ as shown in Fig.~\ref{fig5}, where white nodes are non-observation nodes (but can be observed), while the remaining nodes are observation nodes. For each layer $i,~i\in\llbracket1,l \rrbracket$, there are at most $K$ nodes and exactly one white node ($1+lK\geq n$). In this case, it need at least $1+\left\lfloor \frac{(K-1)(n-1)}{K} \right\rfloor$ observation nodes.

We construct the desired network by taking $N$ disjoint copies of the network shown in Fig.~\ref{fig5}, where the path lengths in these copies are chosen so that the total number of nodes is $n$.
Then, the proposition clearly holds.
\qed

\begin{proposition}
There exists a Leaky-ReLU network with indegree $K$ at every node that is observable from $K$ observation nodes.
\end{proposition}
(Proof)
Construct a Leaky-ReLU network as follows:
\begin{eqnarray*}
x_{Ki+1}(t+1) & = & \rho(a_{Ki+1,K(i+1)+1}x_{K(i+1)+1}(t)+\cdots+a_{Ki+1,K(i+1)+K}x_{K(i+1)+K}(t)+b_{Ki+1}),\\
x_{Ki+2}(t+1) & = & \rho(a_{Ki+2,K(i+1)+1}x_{K(i+1)+1}(t)+\cdots+a_{Ki+2,K(i+1)+K}x_{K(i+1)+K}(t)+b_{Ki+2}),\\
&\ldots&\\
x_{Ki+K}(t+1) & = & \rho(a_{Ki+K,K(i+1)+1}x_{K(i+1)+1}(t)+\cdots+a_{Ki+K,K(i+1)+K}x_{K(i+1)+K}(t)+b_{Ki+K}),
\end{eqnarray*}
where $i\in\llbracket0,l \rrbracket$ and $\begin{vmatrix}
a_{Ki+1,K(i+1)+1} & a_{Ki+1,K(i+1)+2} & \cdots & a_{Ki+1,K(i+1)+K} \\
a_{Ki+2,K(i+1)+1} & a_{Ki+2,K(i+1)+2} & \cdots & a_{Ki+2,K(i+1)+K} \\
\vdots & \vdots & \ddots & \vdots \\
a_{Ki+K,K(i+1)+1} & a_{Ki+K,K(i+1)+2} & \cdots & a_{Ki+K,K(i+1)+K}
\end{vmatrix}\neq0$. For $i=l$, the indices $K(i+1)$ and $K(i+1)+j$ are identified with $n$ and $j$, respectively, where $j\in\llbracket1,K\rrbracket$.

As in the proof of Proposition \ref{prop:best-upper-k2}, the entire initial state $\xvec(0)$ can be uniquely determined from the observations $x_1(t),\ldots,x_{K}(t),~t\in\llbracket 0,l-1 \rrbracket$. \qed

\begin{figure}[ht]
\centering
\includegraphics[width=5cm]{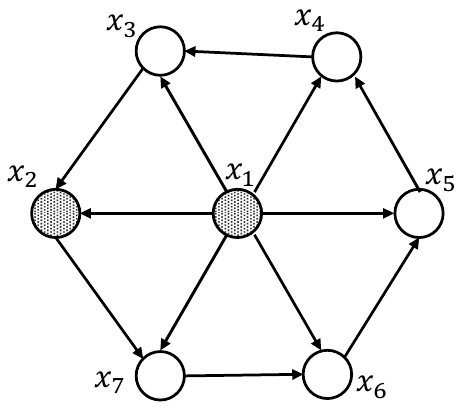}
\caption{Leaky-ReLU networks with star-shaped extended structure.}
\label{fig6}
\end{figure}

Consider the network shown in Fig.~\ref{fig6}. For each $i\in\llbracket3,7\rrbracket$, $x_{i}(t)$ can be uniquely determined from
$x_{1}(t)$ and $x_{i-1}(t+1)$. Applying this relation recursively, the entire initial state $\xvec(0)$ can be uniquely determined from the observations $x_1(t),x_2(t),~t\in\llbracket0,5\rrbracket$.
More generally, we have the following result.
\begin{proposition}
There exists a Leaky-ReLU network for which the minimum number of observation nodes required for observability is $K$. In this network, $K-1$ nodes have outdegree $n-(K-1)$, and the remaining $n-(K-1)$ nodes have outdegree 1 and indegree $K$.
\end{proposition}
(Proof) Construct a Leaky-ReLU network as follows:
\begin{eqnarray*}
x_{K}(t+1) & = & \rho(a_{K,1}x_{1}(t)+\cdots+a_{K,K-1}x_{K-1}(t)+a_{K,K+1}x_{K+1}(t)+b_{K}),\\
x_{K+1}(t+1) & = & \rho(a_{K+1,1}x_{1}(t)+\cdots+a_{K+1,K-1}x_{K-1}(t)+a_{K+1,K+2}x_{K+2}(t)+b_{K+1}),\\
&\ldots&\\
x_{n-1}(t+1) & = & \rho(a_{n-1,1}x_{1}(t)+\cdots+a_{n-1,K-1}x_{K-1}(t)+a_{n-1,n}x_{n}(t)+b_{n-1}),\\
x_{n}(t+1) & = & \rho(a_{n,1}x_{1}(t)+\cdots+a_{n,K-1}x_{K-1}(t)+a_{n,K}x_{K}(t)+b_{n}).
\end{eqnarray*}
Select nodes $x_{1},\ldots,x_{K}$ as observation nodes. From the update equations,
\begin{eqnarray}\label{e1}
x_i(0)=\frac{\rho^{-1}(x_{i-1}(1))-a_{i-1,1}x_{1}(0)-\cdots-a_{i-1,K-1}x_{K-1}(0)-b_{i-1}}{a_{i-1,i}},\quad i\in\llbracket K+1,n \rrbracket.
\end{eqnarray}
Setting $i=K+1$ and $t=0$ in the Eq. (\ref{e1}) uniquely determines $x_{K+1}(0)$ from the observations $x_{1}(0),\ldots,x_{K-1}(0),x_{K}(1)$. To recover $x_{K+2}(0)$, we first use the same relation at $t=1$ to determine $x_{K+1}(1)$, and then apply it at $t=0$ with $i=K+2$. Repeating this process, $x_{K+r}(0)$ can be uniquely determined from the observations $x_{K}(r)$ and $x_1(t),\ldots,x_{K-1}(t),~t\in\llbracket 0,r-1\rrbracket$ for every $r\in\llbracket 1,n-K\rrbracket$. Thus, the entire initial state $\xvec(0)$ can be uniquely determined from
$x_1(t),\ldots,x_K(t),~t\in\llbracket 0,n-K\rrbracket$.
\qed

\begin{proposition}
There exists a Leaky-ReLU network with maximum indegree $K$ that is observable from a single observation node.
\label{prop:best-case-k}
\end{proposition}
(Proof)
Construct a Leaky-ReLU network with maximum indegree $K$ as follows:
\begin{eqnarray*}
x_{1}(t+1) & = & \rho(a_{1,2}x_{2}(t)+b_{1}),\\
x_{2}(t+1) & = & \rho(a_{2,1}x_{1}(t)+a_{2,3}x_{3}(t)+b_{2}),\\
x_{3}(t+1) & = & \rho(a_{3,1}x_{1}(t)+a_{3,2}x_{2}(t)+a_{3,4}x_{4}(t)+b_{3}),\\
&\ldots&\\
x_{K}(t+1) & = & \rho(a_{K,1}x_{1}(t)+\cdots+a_{K,K-1}x_{K-1}(t)+a_{K,K+1}x_{K+1}(t)+b_{K}),\\
x_{K+1}(t+1) & = & \rho(a_{K+1,2}x_{2}(t)+\cdots+a_{K+1,K}x_{K}(t)+a_{K+1,K+2}x_{K+2}(t)+b_{K+1}),\\
&\ldots&\\
x_{n-1}(t+1) & = & \rho(a_{n-1,n-K}x_{n-K}(t)+\cdots+a_{n-1,n-2}x_{n-2}(t)+a_{n-1,n}x_{n}(t)+b_{n-1}),\\
x_{n}(t+1) & = & \rho(a_{n,n-K+1}x_{n-K+1}(t)+\cdots+a_{n,n-1}x_{n-1}(t)+a_{n,1}x_{1}(t)+b_{n}).
\end{eqnarray*}
As in the proof of Proposition \ref{prop:best-case}, the entire initial state $\xvec(0)$ can be uniquely determined from the observations $x_1(t),~t\in\llbracket 0,n-1\rrbracket$.
\qed

As for ReLU networks, the following proposition shows that the worst-case lower bound is quite large ($n$), which also means that the general upper bound is $n$. Therefore, the results in this section suggest that Leaky-ReLU networks are much easier to observe than ReLU networks with respect to the number of observation nodes.

\begin{proposition}
Suppose that weight of every edge in an ReLU network is positive.
Then, the number of observation nodes is always $n$.
\label{prop:relu-worst-lower}
\end{proposition}
(Proof)
Suppose that all elements of $\xvec(0)$ are negative.
Then, clearly we have $\xvec(t)=\zerovec$ for all $t\geq1$.
It means that we cannot get any meaningful information on $\xvec(0)$
from $\xvec(t)$ with $t\geq1$.
Therefore, all nodes must be observation nodes.
\qed
\section{Optimal Bounds for the Minimum Number of Control Nodes}\label{sec:con}
In this section, we characterize the minimum number of control nodes required for controllability. Let $K$ denote the maximum outdegree. We first consider the case $K=2$; that is, each node has at most two outgoing edges, and at least one node has exactly two outgoing edges.

Based on Fig.~\ref{fig:basic} in Section \ref{sec:obs}, we obtain the following two corresponding controllability examples.
\begin{figure}[ht]
\centering
\includegraphics[width=8cm]{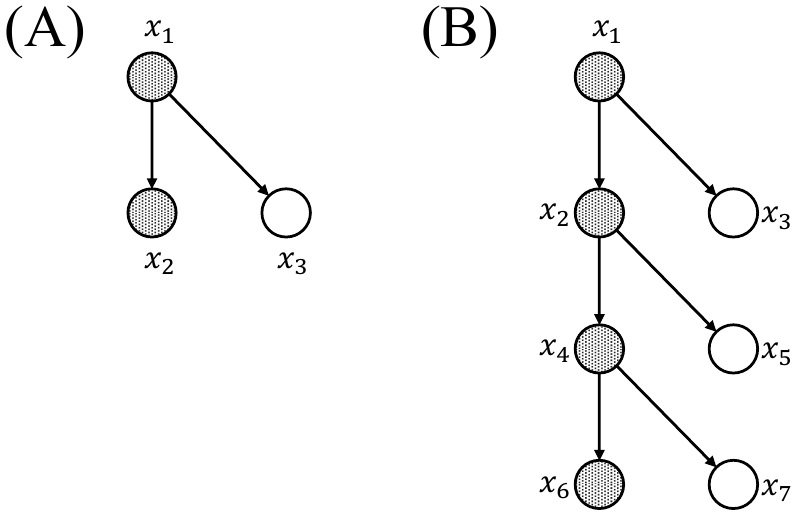}
\caption{Leaky-ReLU networks for explaining the basic idea.
By choosing gray nodes as control nodes,
the system becomes controllable.}
\label{fig7}
\end{figure}

In Fig.~\ref{fig7}(A), let
\begin{eqnarray*}
x_2(t+1) & = & \rho(a_{2,1} x_1(t) + b_2),\\
x_3(t+1) & = & \rho(a_{3,1} x_1(t) + b_3).
\end{eqnarray*}
If $x_{1}$ and $x_{2}$ are selected as control nodes, then $x_{3}(t)$ can be assigned an arbitrary value by suitably controlling $x_{1}(t-1)$. Therefore, the network can be driven from any initial state $\xvec(0)$ to any target state $\xvec^{*}$ at time $t=2$ by controlling $x_{1}(t)$ and $x_{2}(t)$ at $t=1,2$. Specifically, for any $\xvec^{*}=(x_{1}^{*}, x_{2}^{*}, x_{3}^{*})$, choosing $x_{1}(1)=(\rho^{-1}(x_{3}^{*})-b_{3})/{a_{3,1}}$, $x_{1}(2)=x_{1}^{*}$, and $x_{2}(2)=x_{2}^{*}$ yields $\xvec(2)=\xvec^{*}$.

Similarly, in Fig.~\ref{fig7}(B), it is easy to see that the network can be driven from any initial state $\xvec(0)$ to any target state $\xvec^{*}$ at time $t=2$ by controlling $x_{1}(t)$, $x_{2}(t)$, $x_{4}(t)$, and $x_{6}(t)$ at $t=1,2$.

Together with the duality between observability and controllability, the above examples show that for each observable network discussed in the observability part, its controllability counterpart can be obtained by reversing all edge directions and selecting the original observation nodes as control nodes. The resulting edge-reversed network is guaranteed to be controllable.
Let $G'(V,E')$ denote the edge-reversed graph of $G(V,E)$, where $E'=\{(x_{i},x_{j})|(x_{j},x_{i})\in E\}$.

Therefore, the control-node selection problem for the original network can be converted into an observation-node selection problem on its edge-reversed network. Specifically, after applying the procedure $SelObsNodesK2(G(V,E))$ to the edge-reversed network, the selected nodes are taken as the control nodes of the original network. This leads directly to the following results.
\begin{theorem}
For any Leaky-ReLU network with maximum outdegree $K=2$, the minimum number of control nodes required for controllability is at most $N+\lfloor \frac{n-N}{2} \rfloor$. Furthermore, there exists a Leaky-ReLU network with maximum outdegree 2 that need at least $N+\lfloor \frac{n-N}{2} \rfloor$ control nodes. Here, $N$ is the number of top SCCs in its edge-reversed network.
\end{theorem}
\begin{proposition}
There exists a Leaky-ReLU network with outdegree 2 at every node that is controllable with two control nodes.
\end{proposition}
(Proof) Construct a Leaky-ReLU network as follows:
\begin{eqnarray*}
x_{2i-1}(t+1) & = & \rho(a_{2i-1,2(i-1)-1} x_{2(i-1)-1}(t) + a_{2i-1,2(i-1)} x_{2(i-1)}(t) + b_{2i-1}),\\
x_{2i}(t+1) & = & \rho(a_{2i,2(i-1)-1} x_{2(i-1)-1}(t) + a_{2i,2(i-1)} x_{2(i-1)}(t) + b_{2i}),
\end{eqnarray*}
where $i\in\llbracket 1,\frac{n}{2}\rrbracket$, $n$ is even, and
$\begin{vmatrix}
a_{2i-1,2(i-1)-1} & a_{2i-1,2(i-1)} \\
a_{2i,2(i-1)-1} & a_{2i,2(i-1)}
\end{vmatrix}\neq0$.
For $i=1$, the indices $2(i-1)-1$ and $2(i-1)$ are identified with $n-1$ and $n$, respectively.

By setting $x_{2(i-1)-1}(t)$ and $x_{2(i-1)}(t)$ as follows:
\begin{eqnarray*}
\begin{bmatrix}
x_{2(i-1)-1}(t) \\
x_{2(i-1)}(t)
\end{bmatrix}=
\begin{bmatrix}
a_{2i-1,2(i-1)-1} &  a_{2i-1,2(i-1)}\\
a_{2i,2(i-1)-1} & a_{2i,2(i-1)}
\end{bmatrix}^{-1}
\begin{bmatrix}
\rho^{-1}(x_{2i-1}^{*}) - b_{2i-1}\\
\rho^{-1}(x_{2i}^{*}) - b_{2i}
\end{bmatrix},
\end{eqnarray*}
the pair $(x_{2i-1}(t+1),x_{2i}(t+1))$ is driven to the prescribed value $(x_{2i-1}^{*},x_{2i}^{*})$.
Consequently, $x_3(\frac{n}{2})$ and $x_4(\frac{n}{2})$ can be driven to their prescribed values by appropriately choosing $x_1(\frac{n}{2}-1)$ and $x_2(\frac{n}{2}-1)$. Repeated application of the above relation shows that $x_5(\frac{n}{2})$ and $x_6(\frac{n}{2})$ can be driven to their prescribed values by choosing $x_1(\frac{n}{2}-2)$ and $x_2(\frac{n}{2}-2)$. More generally, for each $r\in\llbracket 1,\frac{n}{2}-1\rrbracket$, the pair $(x_{2r+1}(\frac{n}{2}),x_{2r+2}(\frac{n}{2}))$
can be driven to its prescribed value by appropriately choosing
$x_1(\frac{n}{2}-r)$ and $x_2(\frac{n}{2}-r)$.
Thus, by controlling nodes $x_{1}$ and $x_{2}$ over $t\in\llbracket 1,\frac{n}{2}\rrbracket$, the network can be driven from any initial state $\xvec(0)$ to any target state $\xvec^{*}$ at time $t=\frac{n}{2}$.
\qed

It is straightforward to see that the results can be extended to the case $K>2$.
\begin{theorem}
For any Leaky-ReLU networks with maximum outdegree $K$, the minimum number of control nodes required for controllability is at most $N+\left\lfloor \frac{(K-1)(n-N)}{K} \right\rfloor$. Furthermore, there exists a Leaky-ReLU network with maximum outdegree $K$ that need at least $N+\left\lfloor \frac{(K-1)(n-N)}{K} \right\rfloor$ control nodes. Here, $N$ is the number of top SCCs in its edge-reversed network.
\end{theorem}
\begin{proposition}
There exists a Leaky-ReLU network with outdegree $K$ at every node that is controllable with $K$ control nodes.
\end{proposition}
\begin{proposition}
There exists a Leaky-ReLU network for which the minimum number of control nodes required for controllability is $K$. In this network, $K-1$ nodes have indegree $n-(K-1)$, and the remaining $n-(K-1)$ nodes have indegree 1 and outdegree $K$.
\end{proposition}

\begin{proposition}
There exists a Leaky-ReLU network with maximum outdegree $K$ that is controllable with a single control node.
\end{proposition}

\section{Extension to Networks with Invertible Activation Functions}
In the previous sections, we focused on Leaky-ReLU networks. We now show that the specific form of the Leaky-ReLU function is not essential. Instead, the key property used in the previous arguments is the invertibility of the activation functions.

Consider the following network
\begin{eqnarray*}
x_i(t+1) & = & \rho_i\left(\sum_{j=1}^{d^{-}(x_{i})}a_{i,j}x_{i_j}(t)+b_i\right),
\quad i\in\llbracket 1,n\rrbracket,
\end{eqnarray*}
where each activation function $\rho_i$ is injective. In other words, $\rho_i^{-1}$ is well defined for every $i$.

For the observability results, it is enough to assume that $\rho_i$ is injective. In this case, $\sum_{j=1}^{d^{-}(x_{i})}a_{i,j}x_{i_j}(t)+b_i$ is uniquely determined from $x_i(t+1)$. Hence, the bounds on the number of observation nodes also hold in the following cases:
\begin{itemize}
  \item $\rho_i$ is a strictly increasing piecewise-linear function for each $i$.
  \item $\rho_i$ is a strictly decreasing piecewise-linear function for each $i$.
  \item $\rho_i$ is a Leaky-ReLU function with a node-dependent slope $\alpha_i\in(0,1)$.
  \item Some $\rho_i$ are strictly increasing piecewise-linear functions, while others are strictly decreasing piecewise-linear functions.
\end{itemize}

In contrast, for the controllability results with arbitrary target states in $\mathbb R^n$, we further require each activation function $\rho_i$ to be bijective. For example, $\rho(s)=\tanh(s)$ is injective but not surjective, since its range is $(-1,1)$. Thus, arbitrary target states in $\mathbb{R}^n$ cannot in general be reached, and the bounds on the minimum number of control nodes do not directly extend to such activation functions.
\section{Simulation Results}
In this section, we present numerical experiments illustrating the minimum number of observation nodes required for Leaky-ReLU network observability.
For each triple $(n,K,N)$, where $n$ is the number of nodes, $K$ is the maximum indegree, and $N$ is the number of top SCCs, we randomly generated 30 Leaky-ReLU networks. For each generated network, the exact minimum number of observation nodes is obtained by exhaustive enumeration over all node subsets. The computed exact values are then compared with the theoretical general lower bound and general upper bound.

\begin{figure*}\centering
\begin{subfigure}[b]{0.3\textwidth}
  \centering
  \includegraphics[width=\textwidth]{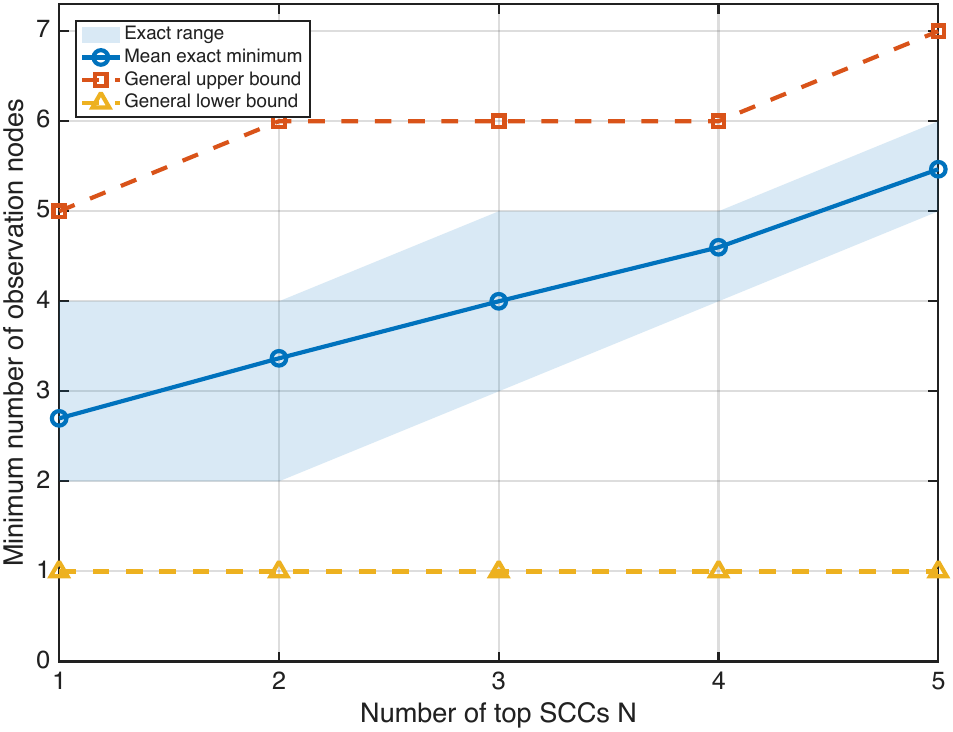}
  \caption{}
\end{subfigure}
\begin{subfigure}[b]{0.3\textwidth}
  \centering
  \includegraphics[width=\textwidth]{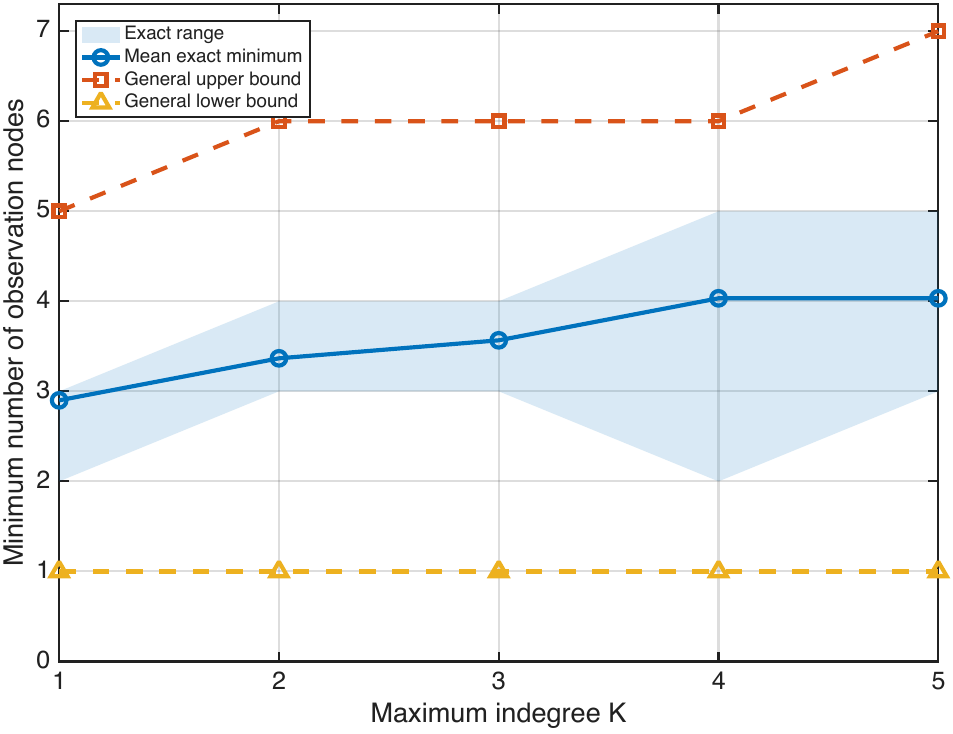}
  \caption{}
\end{subfigure}
\begin{subfigure}[b]{0.3\textwidth}
  \centering
  \includegraphics[width=\textwidth]{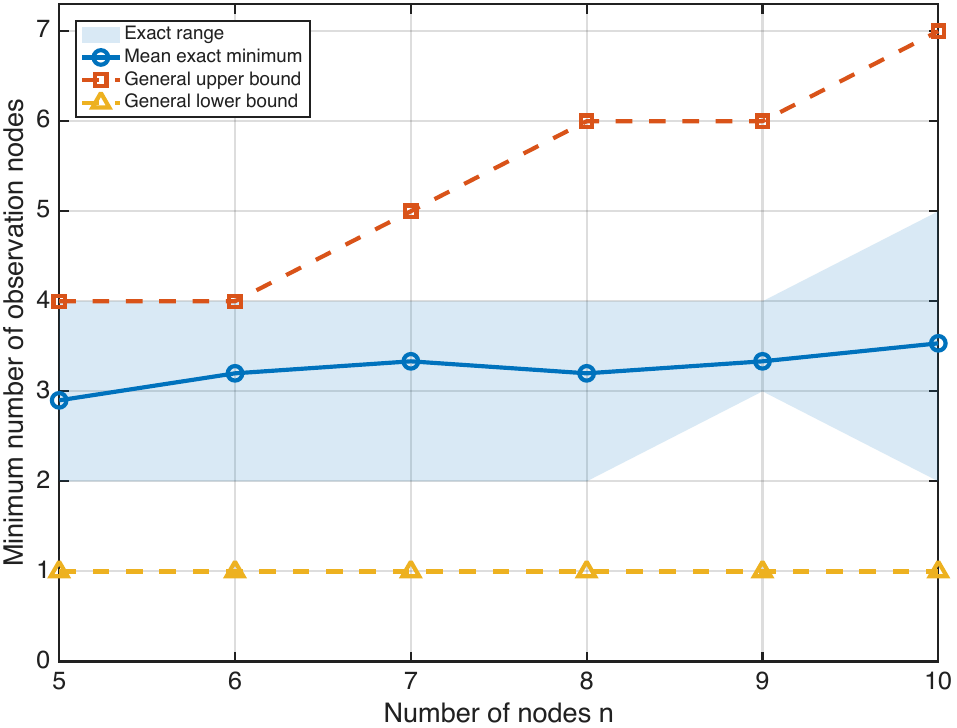}
  \caption{}
\end{subfigure}
\caption{Comparison between the exact minimum number of observation nodes and the derived bounds.
The blue solid line represents the average of the exact minimum values on randomly generated networks, while the shaded region indicates the range of these exact minimum values.
The red dashed line and the yellow dashed line represent the worst-case lower bound and the general upper bound, respectively.
(a) Fixed $n=8$ and $K=3$, with $N$ varying.
(b) Fixed $n=8$ and $N=2$, with $K$ varying.
(c) Fixed $K=3$ and $N=2$, with $n$ varying.}\label{fig8}
\end{figure*}

The results obtained by MATLAB simulations on a MacBook Pro (Apple
M4 Pro, 24 GB RAM) are summarized in Fig. \ref{fig8}, which indicates that the derived general upper bound is valid for the tested instances, while the randomly generated networks are much easier to observe than the worst-case constructions.
Moreover, although the general lower bound of one is shown to be tight, the exact minimum number of observation nodes is at least two for all randomly generated networks tested. This indicates that the networks attaining the general lower bound are special constructions, whereas randomly generated networks typically require more observation nodes.
\section{Conclusions}
This paper investigated minimum-node observability and controllability of Leaky-ReLU networks. For networks with node indegrees bounded by $K$, a graph-theoretic analysis yielded a procedure for selecting observation nodes and established an upper bound ($N+\lfloor \frac{(K-1)(n-N)}{K} \rfloor$) on the minimum number of observation nodes required for observability. We then constructed a family of networks attaining this bound, thereby determining the exact worst-case observation-node requirement. We also identified network structures that are observable from a single node over a finite horizon. Since at least one observation node is
necessary, this construction establishes the exact best-case requirement of one.

By exploiting the observability--controllability duality, we obtained the corresponding best- and worst-case results for the minimum number of control nodes under the associated degree constraints. A comparison with ReLU networks further showed how replacing the zero negative slope of ReLU with a nonzero slope changes the observation-node requirements. Finally, the
analysis is not specific to Leaky-ReLU networks: the observability results extend to networks with componentwise injective activation functions, whereas the controllability results extend to networks with componentwise bijective activation functions.

\end{document}